\documentclass[12pt] {article}
\def\be{\begin{equation}}
\def\ee{\end{equation}}
\def\bea{\begin{eqnarray}}
\def\eea{\end{eqnarray}}
\def\pd{\partial}

\begin{document}
\begin{titlepage}

\title{The Multi-field Complex Bateman  Equation}

\author{D.B. Fairlie$\footnote{e-mail: david.fairlie@durham.ac.uk}$\\
\quad\\
{ Department of Mathematical Sciences}\\
{ University of Durham, Durham DH1 3LE}}
\maketitle

\begin{abstract}
The multi-field generalisation of the Bateman equation arises from considerations of the continuation of String and Brane equations to the case where the base space is of higher dimension than the target space. The complex 
extension of this equation possesses a remarkably large invariance group, and 
admits a very simple implicit form for its general solution, in addition to the special case of holomorphic and anti-holomorphic explicit solutions. A class of inequivalent Lagrangians for this equation is discovered. 
\end{abstract}
\end{titlepage}

\section{Introduction}
In a recent attempt to extend the inference from the point particle Lagrangian to the Klein Gordon Lagrangian, the concept of of a Companion Lagrangian associated with  Strings and Branes was introduced \cite{baker1,baker2}. 
Another way of thinking about it is to consider the continuation of the
String and Brane Lagrangians to the situation where the base space is of higher dimensionality than the target space.
The simplest example, of the point particle in two dimensions  gives rise to the well known Bateman equation; 
\be
\left(\frac{\pd \phi}{\pd x_1}\right)^2\frac{\pd^2 \phi}{\pd x_2^2}+
\left(\frac{\pd \phi}{\pd x_2}\right)^2\frac{\pd^2 \phi}{\pd x_1^2}-
2\left(\frac{\pd \phi}{\pd x_1}\right)\left(\frac{\pd \phi}{\pd x_2}\right)\frac{\pd^2 \phi}
{\pd x_1\pd x_2}\,=\,0.\label{batman}
\ee
This equation and its multi-field generalisations, which are the equations of motion for the higher dimensional Companion Lagragians have been extensively
studied, and are known to be integrable\cite{gov1,gov2,gov3,gov4}.  Recently
a complex version of the Bateman equation, and its generalisation to higher space dimensions has been shown to possess a remarkably elegant implicit solution by Leznov\cite{lez1,lez2}. One of the puzzles with this equation is that there is no evident Lagrangian formalism for it.  The purpose of this paper is to redress this omission and construct a Lagrangian for the Complex Bateman equation and its multi-field generalisation, and to find the general solution of the latter. 
 The equations 
presented here are no worse than second order, are covariant and possess reparametrisation invariance. Their solutions, though implicit, are contained in an elegant series of equations. Among the class of explicit solutions there exists a simple generalisation of the  holomorphic plus anti-holomorphic function solution of the two dimensional Laplace equation.

\section{The Complex Bateman equation}
The Complex Bateman equation
\be
\frac{\pd\phi}{\pd x_1}\frac{\pd\phi}{\pd y_1}\frac{\pd^2\phi}{\pd x_2\pd y_2}+\frac{\pd\phi}{\pd x_2}\frac{\pd\phi}{\pd y_2}\frac{\pd^2\phi}{\pd x_1\pd y_1}-\frac{\pd\phi}{\pd x_1}\frac{\pd\phi}{\pd y_2}\frac{\pd^2\phi}{\pd x_2\pd y_1}-\frac{\pd\phi}{\pd x_2}\frac{\pd\phi}{\pd y_1}\frac{\pd^2\phi}{\pd x_1\pd y_2}\,=\,0,\label{one}
\ee
where $(y_1,y_2)$ may be thought of as complex conjugates of  $(x_1, x_2)$, is an
equation with a huge invariance group. It is covariant; if  $\phi$ is a solution, so is any differentiable function of $\phi$. It is also reparametrisation invariant under both reparametrisations of the variables  $(x_1,x_2)$ and  $(y_1, y_2)$. It
may be solved completely in terms of two arbitrary functions $H(\phi,y_1,y_2)$ and $K(\phi,x_1,x_2)$  which are constrained to be equal\cite{lez1};
\be H(\phi,y_1,y_2) \,=\,K(\phi,x_1,x_2).\label{lez}\ee
The solution of this equation for $\phi$ gives an implicit solution to (\ref{one}), and the form of the solution reflects the statements made about 
the invariance group of the equation.
An equivalent form of the solution occurs much earlier in the literature, \cite{chaundy}. The elimination of $\psi$ from the pair of equations for two arbitrary functions of four variables;  
\be F(\phi,\psi,x_1,x_2) \,=\,0;\ \ \ \ G(\phi,\psi, y_1,y_2)\,=\,0,\label{two}
\ee
gives an implicit realisation of the general solution.
Each member of the pair
\bea 
\frac{\pd u}{\pd y_1}&=&v\frac{\pd u}{\pd y_2}\label{five}\\
\frac{\pd v}{\pd x_1}&=& u\frac{\pd v}{\pd x_2}\label{six}
\eea
when
\[ u\,=\, \frac{\frac{\pd\phi}{\pd x_1}}{\frac{\pd\phi}{\pd x_2}};\ \ \ 
v\,=\, \frac{\frac{\pd\phi}{\pd y_1}}{\frac{\pd\phi}{\pd y_2}}\]
separately implies (\ref{one}). Furthermore, these  last pair of equations follow from (\ref{five},\ref{six}) under the hypothesis that $u$ is a function of three variables;  $(\phi,x_1,x_2)$ and $v\,=\, v(\phi,y_1,y_2)$. The resulting constraints arising from the two different expressions for  $u,\ v$, namely
\[ u(\phi,x_1,x_2)\,=\, \frac{\frac{\pd\phi}{\pd x_1}}{\frac{\pd\phi}{\pd x_2}};\ \ \ 
 v(\phi,y_1,y_2)\,=\, \frac{\frac{\pd\phi}{\pd y_1}}{\frac{\pd\phi}{\pd y_2}}\]
both separately imply the complex Bateman equations by themselves.
  The clue as to how to discover a Lagrangian 
lies in the second formulation of the solution;
It turns out that the introduction of the second function is crucial and a possible Lagrangian is (reverting to a notation in which subscripts denote derivatives)
\be
{\cal L} = \left(\psi_{x_1}\phi_{x_2}-\psi_{x_2}\phi_{x_1}\right)\left(\frac{\phi_{y_1}}
{\phi_{y_2}}\right) \,+\, {\rm c.c.}.\label{lag1}
\ee
Note that this Lagrangian is singular.
The equation of motion which comes from variations in the field $\psi$ is simply
\bea
&&\frac{\pd}{\pd x_1}\left(\phi_{x_2}\frac{\phi_{y_1}}{\phi_{y_2}}\right)\,-\,
\frac{\pd}{\pd x_2}\left(\phi_{x_1}\frac{\phi_{y_1}}{\phi_{y_2}}\right)\\
\label{eqm}
&=&\frac{1}{\phi_{y_2}^2}\left(\phi_{x_2}\phi_{y_2}\phi_{x_1y_1}-\phi_{x_2}\phi_{y_1}\phi_{x_1y_2}-\phi_{x_1}\phi_{y_2}\phi_{x_2y_1}+\phi_{x_1}\phi_{y_1}\phi_{x_2y_2}\right)\,=\,0\nonumber.
\eea
This is just the equation of motion (\ref{one}). When variations in the field $\phi$ are considered, and (\ref{one}) is imposed, the resulting equation implies that  $\psi$ may be taken to be any differentiable function of $\phi$.
Thus the equations of motion are (\ref{one}) for both $\phi$ and $\psi$ since the equation of motion is also satisfied by any function of $\phi$.
Just as the real Bateman equation possesses an infinite number of inequivalent Lagrangians, dependent only upon the first derivatives of the field \cite{gov1},
so does the complex Bateman equation. Any Lagrangian of the form
\be
{\cal L} = \left(\psi_{x_1}\phi_{x_2}-\psi_{x_2}\phi_{x_1}\right){\cal K}(\phi_{y_1},\phi_{y_2})\,+\, {\rm c.c.}
 \label{lag2}
\ee
will do where ${\cal K}(\phi_{y_1},\phi_{y_2})$ is of weight zero in its arguments, i.e. 
\[ \phi_{y_1}\frac{\pd{\cal K}}{\pd\phi_{y_1}}\,+\,\phi_{y_2}\frac{\pd{\cal K}}{\pd\phi_{y_2}}\,=\,0.\]
Note further that under a combined reparametrisation of the pairs of independent variables, the first term in (\ref{lag2}) simply transforms as the Jacobian of the transformation of $(x_1, x_2)$.
\subsection{Explicit Solutions}
An evident and possibly useful explicit solution of the equation of motion is to take
\be
\phi\,=\, f(x_1,x_2)\,+\, g(y_1,y_2),\label{holo}
\ee
where $f$ and $g$ are arbitrary `holomorphic'  and `anti-holomorphic' functions.
These solutions come from the general formalism (\ref{lez}) by setting, for example $K\,=\, \phi- f(x_1, x_2)\ \ H\,=\, g(y_1,y_2).$
If either $f$ or $g$ vanishes, if one is not too fastidiously rigorous  the action is zero.
Inserting as a trial solution $ \phi\, = \, ax_1y_1 + bx_1y_2 + cx_2y_1 + dx_2y_2$ the right hand side of (\ref{one}) reproduces
\[ (ad-bc)(ax_1y_1 + bx_1y_2 + cx_2y_1 + dx_2y_2)\]
so any function of this  quadratic will be a solution if $ad-bc\,=\,0.$

\section{Multi-field Generalisation}
 The extension to the complex form of the multi-field Bateman follows the pattern of that for two fields $\phi$ and $\psi$ dependent upon six co-ordinates three designated by $(x_1,x_2,x_3)$ and three others by $(y_1,y_2,y_3)$ which may be regarded as their complex conjugates.
The equations of motion are
\be
\det\left|\begin{array}{ccccc}
0&0&\phi_{y_1}&\phi_{y_2}& \phi_{y_3} \\
0&0&\psi_{y_1}&\psi_{y_2}& \psi_{y_3} \\
\phi_{x_1}&\psi_{x_1}&\phi_{x_1y_1}&\phi_{x_1y_2}&\phi_{x_1y_3}\\
\phi_{x_2}&\psi_{x_2}&\phi_{x_2y_1}&\phi_{x_2y_2}&\phi_{x_2y_3}\\
\phi_{x_3}&\psi_{x_3}&\phi_{x_3y_1}&\phi_{x_3y_2}&\phi_{x_3y_3}\end{array}
\right|\,=\,0,\label{threetwo}
\ee
together with another equation in which second derivatives of $\phi$ are replaced by second derivatives of $\psi$. This equation exhibits directly the properties of covariance (any pair of functions of $\phi,\ \psi$ are solutions
if $\phi$ and $\psi$ are) as well as reparametrisation invariance in redefinitions of $(x_1, x_2, x_3)$ and also of $(y_1,y_2, y_3)$
The surprising result is that the general solution to these equations is obtained by taking four arbitrary functions of five variables
set equal in pairs;
\bea
U^1(\phi,\psi,x_1,x_2,x_3)&=&V^1(\phi,\psi,y_1,y_2,y_3),\label{pair1}\\ U^2(\phi,\psi,x_1,x_2,x_3)&=& V^2(\phi,\psi,y_1,y_2,y_3).\label{pair2} 
\eea
This claim may be verified by differentiating  the first equality twice; first with respect to $x_i$, then with respect to $y_j$, multiplying by the Jacobians
$\displaystyle {J_i\,=\, \epsilon_{ikl}\phi_{x_k}\psi_{x_l}}$ and $\displaystyle{\bar J_j\,=\, \epsilon_{jkl}\phi_{y_k}\psi_{y_l}}$ and summing over $i,\ j$.
All terms without a second derivative of either $\phi$ or $\psi$ disappear and
we obtain
\be
\left(\frac{\pd U^1}{\pd \phi}-\frac{\pd V^1}{\pd \phi}\right)J_i\bar J_j\phi_{x_iy_j}+\left(\frac{\pd U^1}{\pd \psi}-\frac{\pd V^1}{\pd \psi}\right)J_i\bar J_j\psi_{x_iy_j}\,=\,0.\label{eqtwo}
\ee
This gives a linear combination of the two equations for $\phi,\ \psi$. In the generic case the second equality (\ref{pair2}) gives a linearly independent combination of the same equations, 
\be
\sum_{i,j}J_i\bar J_j\phi_{x_iy_j}\,=\,0;\ \ \ \sum_{i,j}J_i\bar J_j\phi_{x_iy_j}\,=\,0,\label{same}
\ee 
which are just another way of writing (\ref{threetwo}) and the corresponding equation with second order $\psi$ derivatives. 
The arbitrariness in the choice of functions corresponds to the arbitrariness in the general solution. 
The equations may also be expressed in a first order form of equations of hydrodynamic type by introducing the Jacobians
\bea
u_1&=&\frac{\phi_{x_1}\psi_{x_2}-\phi_{x_2}\psi_{x_1}}{\phi_{x_2}\psi_{x_3}-\phi_{x_3}\psi_{x_2}}\ \ \ u_2\,=\,\frac{\phi_{x_3}\psi_{x_1}-\phi_{x_1}\psi_{x_3}}{\phi_{x_2}\psi_{x_3}-\phi_{x_3}\psi_{x_2}},\label{soln1}\\
v_1&=&\frac{\phi_{y_1}\psi_{y_2}-\phi_{y_2}\psi_{y_1}}{\phi_{y_2}\psi_{y_3}-\phi_{y_3}\psi_{y_2}}\ \ \ v_2\,=\,\frac{\phi_{y_3}\psi_{y_1}-\phi_{y_1}\psi_{y_3}}{\phi_{y_2}\psi_{y_3}-\phi_{y_3}\psi_{y_2}}.\label{soln2}
\eea
The equations take the form
\bea
\frac{\pd u_1}{\pd y_1}&+& v_1\frac{\pd u_1}{\pd y_3}\,+\, v_2\frac{\pd u_1}{\pd y_2}\,=\,0,\label{hydro1}\\
\frac{\pd u_2}{\pd y_1}&+& v_1\frac{\pd u_2}{\pd y_3}\,+\, v_2\frac{\pd u_2}{\pd y_2}\,=\,0,\label{hydro2}\\
\frac{\pd v_1}{\pd x_1}&+& u_1\frac{\pd v_1}{\pd x_3}\,+\, u_2\frac{\pd v_1}{\pd x_2}\,=\,0,\label{hydro3}\\
\frac{\pd v_2}{\pd x_1}&+& u_1\frac{\pd v_2}{\pd x_3}\,+\, u_2\frac{\pd v_2}{\pd x_2}\,=\,0.\label{hydro4}
\eea
Compare the situation for the real case in \cite{baker1}. Here there are  just two equations;
\bea
\frac{\pd u_1}{\pd x_1}+u_1\frac{\pd u_1}{\pd x_3}+u_2\frac{\pd u_1}{\pd x_2}&=&0,
\nonumber\\
\frac{\pd u_2}{\pd x_1}+u_1\frac{\pd u_2}{\pd x_3}+u_2\frac{\pd u_2}{\pd x_2}&=&0.
\label{eex}
\eea
Under the assumption that both $u_1$ and $u_2$ depend upon a pair of functions
$\phi,\psi$, the Companion Equations for Strings in three dimensions results;
if however  $u_1$ and $u_2$ depend only upon a single function $\phi(x_1,x_2,x_3)$ then the outcome is the so-called Universal Field Equation \cite{gov1} for $\phi$ in three  \hfill\break dimensions;

\[
\det\left|\begin{array}{cccc}
0&\phi_{x_1}&\phi_{x_2}& \phi_{x_3} \\
\phi_{x_1}&\phi_{x_1x_1}&\phi_{x_1x_2}&\phi_{x_1x_3}\\
\phi_{x_2}&\phi_{x_2x_1}&\phi_{x_2x_2}&\phi_{x_2x_3}\\
\phi_{x_3}&\phi_{x_3x_1}&\phi_{x_3x_2}&\phi_{x_3x_3}\end{array}
\right|\,=\,0.\]

 This assertion is proved in generality in \cite{integrable}. In the complex case, it is shown in \cite{lez2}
that the equations (extended to arbitrary dimension) reproduce the complex form
 of the Bateman equation in higher dimensions; if on the other hand the functions $u_1,u_2,v_1,v_2$ are regarded as dependent upon five variables, $u_i\,=\,u_i(\phi,\psi,x_1,x_2,x_3)$  and $v_i\,=\,v_i(\phi,\psi,y_1,y_2,y_3)$ then the multi-field Bateman equation (\ref{threetwo}) results. To see this notice that that under this hypothesis (\ref{hydro1}) and  (\ref{hydro2}) yield 
\bea
\frac{\pd u_1}{\pd \phi}\left(\phi_{y_1} +v_1\phi_{y_3}+ v_2\phi_{y_2}\right)+\frac{\pd u_1}{\pd \psi}\left(\psi_{y_1} +v_1\psi_{y_3}+ v_2\psi_{y_2}\right)&=&0,\nonumber\\
\frac{\pd u_2}{\pd \phi}\left(\phi_{y_1} +v_1\phi_{y_3}+ v_2\phi_{y_2}\right)+\frac{\pd u_2}{\pd \psi}\left(\psi_{y_1} +v_1\psi_{y_3}+ v_2\psi_{y_2}\right)&=&0.\nonumber
\eea
The solution of these equations is simply given by the pair of equations
(\ref{soln2}). Similarly the solution of (\ref{hydro3}) and  (\ref{hydro4}) 
gives (\ref{soln1}) and thus the equations of motion are reproduced. 
The consistency of the pair of equations (\ref{soln1}) under the assumption of the functional dependence \hfill\break$u_i\, =\,u_i(\phi(x_j,y_j),\psi(x_j,y_j),x_1,x_2,x_3)$
itself implies the multi-field equations. If instead $u_i\,=\,u_i(\phi(x_j,y_j),x_1,x_2,x_3)$ so there is no dependence upon $\psi$ , and the same  for $v_i$ then (\ref{hydro1}) and (\ref{hydro2}) are the same equation for $\phi$. This equation, together with the three others obtained by partial differentiation with respect to $(y_1,y_2,y_3)$ are linear equations whose eliminant is the complex Bateman equation in three dimensions\cite{lez2}.
Thus it appears that the first order equations are more fundamental, giving rise to 
different possibilities for  second order equations, depending upon the assumptions made about the dependence of the functions $(u_j,\ v_j)$ upon the dependent variables in the second order equations.

\subsection{Multi-field Lagrangian}
The Lagrangian can be constructed along similar lines to that for the single field; 
one choice is
\be
{\cal L}=\frac{\pd(\phi,\ \psi,\ \theta)}{\pd(x_1, x_2, x_3)}v_1\,+\, {\rm c.c.}
\label{lag3}
\ee
Variation with respect to $ \theta$ gives a combination of the equations of motion for $\phi$ and $\psi$; variations with respect to the latter fields yields other linear combinations which together imply  (\ref{threetwo}) 
and that $\theta$ is a function of $\phi$ and $\psi$, in much the same manner as
the single field case.
A more general choice is 
\be
{\cal L}=\frac{\pd(\phi,\ \psi,\ \theta)}{\pd(x_1, x_2, x_3)}{\cal K}(\phi_{y_1},\phi_{y_2},\phi_{y_3},\psi_{y_1},\psi_{y_2},\psi_{y_3}),
\label{lag4}
\ee
where ${\cal K}$ is a homogeneous function,  of weight zero in both sets of derivatives; i.e.
\bea
 \phi_{y_1}\frac{\pd{\cal K}}{\pd\phi_{y_1}}\,+\,\phi_{y_2}\frac{\pd{\cal K}}{\pd\phi_{y_2}}+ \phi_{y_3}\frac{\pd{\cal K}}{\pd\phi_{y_3}}&=&0,\nonumber\\
 \psi_{y_1}\frac{\pd{\cal K}}{\pd\psi_{y_1}}\,+\,\psi_{y_2}\frac{\pd{\cal K}}{\pd\psi_{y_2}}+ \psi_{y_3}\frac{\pd{\cal K}}{\pd\psi_{y_3}}&=&0.\nonumber
\eea
In addition we require
\bea
 \psi_{y_1}\frac{\pd{\cal K}}{\pd\phi_{y_1}}\,+\,\psi_{y_2}\frac{\pd{\cal K}}{\pd\phi_{y_2}}+ \psi_{y_3}\frac{\pd{\cal K}}{\pd\phi_{y_3}}&=&0,\nonumber\\
 \phi_{y_1}\frac{\pd{\cal K}}{\pd\psi_{y_1}}\,+\,\phi_{y_2}\frac{\pd{\cal K}}{\pd\psi_{y_2}}+ \phi_{y_3}\frac{\pd{\cal K}}{\pd\psi_{y_3}}&=&0.\nonumber
\eea
A small Lemma is appropriate here;

{\bf Lemma}:
If ${\cal K}(\phi_{y_j},\psi_{y_j})$
is homogeneous of degree $m$ in the derivatives $\phi_{y_j}$  and of degree $n$
in the derivatives $\psi_{y_j}$, then either $\displaystyle{\sum_j\phi_{y_j}\frac{\pd{\cal K}}{\pd\psi_{y_j}}}$ has the respective homogeneities $m+1,\ n-1$ or else is zero. The latter option should be chosen here.
As in the previous case if both $\phi$ and $\psi$ are able to be split into holomorphic and anti-holomorphic contributions, then this will be an explicit solution of the equations of motion. 
The general pattern becomes clear; the equations of motion for $N$ fields in 
$N+1$ dimensions take the form of  (\ref{threetwo}) and the general solution will be obtained from the solution of 
$N$ equations of the form
\be U^i(\phi^j,\ x_1,\dots,x_{N+1})\,=\, V^i(\phi^j,\ y_1,\dots,y_{N+1}).\label{soln} 
\ee
This claim is proven by a similar argument to that for the case where $N=2$;

The Lagrangian takes a similar form of a Jacobian in one set of derivatives multiplied by a homogeneous function of degree zero in the other. If the dimensions of the base space $D$ are greater than $N+1$, then the equations of motion take the form of a linear combination of $D\choose N+1$ equations of the above structure, with a class of solutions formed from combining those for the basic units as in the real case\cite{baker1}.

\section{Conclusions}
The Complex Bateman equation and its multi-field generalisations may be taken as
a paradigm for the study of systems with covariant solutions and a high degree of reparametrisation invariance. They are a natural extension of the equations of motion which arise from the idea of a Companion Lagrangian for Strings and Branes \cite{baker1,baker2}, and admit an infinite number of inequivalent Lagrangians. They admit explicit solutions which are a linear combination of
holomorphic and anti-holomorphic functions, and their general solution 
is very simply characterised in implicit form in terms of the solution of a set of equalities. Further developments which may be anticipated are the discovery of a larger class of Lagrangians, which will treat the two sets of co-ordinates on a more equal footing.
  A connection is made with a series of first order equations of hydrodynamic type which are more fundamental, as they yield different possibilities for the resulting second order equations, according to the functional dependencies assumed for the `velocities' in these first order equations. This situation calls for further investigation.  
Some consideration has been given to the question of how to connect the appearance of covariance in the real multi-field equations with the usual signal for the presence of a covariant derivative \cite{ueno}, and a similar question may be asked of this complex version. 
\section*{Acknowledgement} The author is indebted to the Leverhulme trust for the award of an Emeritus Fellowship.

\end{document}